\documentclass{article}
\usepackage[utf8]{inputenc}
\usepackage[english]{babel}
\usepackage[pdfusetitle, colorlinks=true, linktoc=page, allcolors=blue]{hyperref}
\usepackage{inconsolata}
\usepackage{graphicx}
\usepackage{subcaption}
\usepackage{listings}
\usepackage{xcolor}
\usepackage{indentfirst}
\usepackage[numbers]{natbib}
\usepackage{amsmath,amssymb, amsthm}
\usepackage{multirow}
\usepackage{caption}
\usepackage{amsmath}

\usepackage{titling}
\usepackage{authblk}
\makeatletter
\def\maketitle{\AB@maketitle}

\makeatother

\usepackage[left=2cm, right=2cm,
top=2cm,bottom=2cm,bindingoffset=0cm]{geometry}
\newtheorem{theorem}{Theorem}

\newtheorem*{corollary*}{corollary}

\begin{document}
\title{Modelling optimal implementation of an arbitrary
\texorpdfstring{$N$}{N}-qubit quantum gate within the generalized
Bloch vectors formalism}
\author[1, *]{\textbf{Elena R. Loubenets}}
\author[1, 2]{\textbf{Sergey Kuznetsov}}

\affil{Department of Applied Mathematics, MIEM, HSE University, Moscow 123458, Russia}
\affil[2]{Steklov Mathematical Institute of Russian Academy of Sciences,
Moscow 119991, Russia}

\date{}

\maketitle

\begin{abstract}
The optimal implementation of quantum gates for closed $N$-qubit systems is one of the key challenges
for practical realization of many quantum information processing tasks. In the present article, based on the generalized Bloch vectors formalism [\emph{J. Phys. A: Math. Theor.} 54, 195301 (2021)] for a finite-dimensional quantum system, we develop a new \emph{general} model for the optimal implementation of quantum gates, which is formulated in terms of the Bloch vectors for the unitary evolution operator and the system Hamiltonians, drift and control, and has the \textcolor{black}{unified mathematical form,  which is standard for the mathematical control theory and applicable} for the implementation of an arbitrary $N$-qubit gate within any closed $N$-qubit system, satisfying the controllability conditions. Within the developed optimal model, the cost functional has both the terminal part and also, the integral part with the special scaling, and this allows us to specify the  quantum optimal control synthesis via solving the two-point boundary value problem (BVP) for the system of ordinary differential equations (ODEs), which can be
explored numerically by any of the known BVP solvers for ODEs. The numerical experiments,  conducted for the implementation within the developed optimal model of a variety of $N=1,2,3$ qubit gates, demonstrate the high accuracy of the model-based results.

{\color{black} \par\medskip
}
\end{abstract}

\section{Introduction}

\begingroup
  \renewcommand{\thefootnote}{\fnsymbol{footnote}}
  \footnotetext[1]{Author to whom the correspondence should be addressed: \textbf{elena.loubenets@hse.ru}}
\endgroup

The key problem for the practical implementation of different quantum applications tasks is finding an optimal action on a quantum system, which, depending on an optimal control goal, either leads a quantum system state to a target quantum state or ensures the optimal execution of a certain quantum operation.

The analytical methods used in quantum control theory \cite{2,1,3,4,5} are based both on general methods \cite{6,6',7,9,10} of the mathematical theory of optimal control -- the Pontryagin maximum principle, the Bellman optimal control principle, the Krotov method, and the methods developed specifically for the optimal control over quantum systems, see \cite{5', 11,12,13} and references therein. 

Similarly, the numerical analysis of quantum optimal control problems involves the use of general numerical optimization algorithms, including various versions of gradient-based and gradient-free algorithms, as well as the numerical optimization algorithms, developed specifically for the optimal quantum control problems. For details, see Section 3 in \cite{1}, Introduction in \cite{13'} and Section 4.1 in \cite{13"} with the list of packages for quantum optimal control. 

Due to the development of large-scale quantum processors, one of the current relevant research areas in the theory of optimal quantum control is the development of models  for the optimal implementation of $N\geq1$ qubit quantum gates. For this purpose, various approaches are applied.  

In the optimal models, proposed in \cite{14,16,18}, the implementation of $N$-qubit quantum gates is based on bringing some initial $N$-qubit state within its evolution under control to the target state which is specified by the action on this initial state of a quantum gate to be generated.
In the frame of these models, a number of the used initial states varies from one in case of the optimal realization of one-qubit gate NOT in \cite{16} to three in case of the optimal implementation of two-qubit gates \cite{18} and also $N$-qubit gates \cite{14}. The latter choice is due to the fact \cite{14} that, in order to find a control optimally implementing a quantum gate via bringing an arbitrary initial $N$-qubit state to the target one due to the terminal cost functional specified by the distance between the final state and this target state, it is suffice to use exactly three specific initial $N$-qubit states constructed in \cite{14}.

Note that each $N$-qubit quantum gate is represented on the Hilbert space $(\mathbb{C}^{2})^{\otimes N}$ by a unitary operator $U_{\text{gate}},$ describing the action $|\psi_{\text{in}}\rangle\to|\psi_{\text{out}}\rangle=U_{\text{gate}}|\psi_{\text{in}}\rangle$ of this gate on an $N$-qubit state $|\psi_{\text{in}}\rangle$ "entering"\space the black box describing this gate schematically on a quantum circuit. Therefore, for the optimal implementation of an $N$-qubit gate $U_{\text{gate}}$ within the unitary evolution of an $N$-qubit system, it is necessary to optimize, based on the chosen cost  functional, the control parameters that bring the unitary evolution operator $U(t)$ of an $N$-qubit system at the final time to the needed target value $U_{\text{gate}}$. 
For $N$-qubit systems with specific Hamiltonians and the gate fidelity as the cost functional, this ”direct” approach to $N$-qubit gates implementation is explored experimentally in several papers, see \cite{19_1, 19_2} and references therein.  
 
In \cite{21,22,19}, the optimal implementation of two-qubit quantum gates is due to bringing to the target value via different terminal cost functionals of the superoperator, describing evolution of an open two-qubit system.

The comparative study 
of different optimal control techniques, used for the implementation of gates NOT and CNOT,  is presented in \cite{19'}.

In the present paper, within the unitary evolution of an $N$-qubit quantum system, we develop analytically and  investigate further numerically a new \emph{general} model \eqref{BFCS_model_statement} for the optimal implementation of an arbitrary $N$-qubit quantum gate. This new model (i) is not related to the specific choice of initial $N$-qubit system states, as it is the case in \cite{14,16,18}, (ii) has the unified mathematical form, \textcolor{black}{which is standard for the mathematical control theory and applicable} for the implementation of an arbitrary $N\geq 1$ qubit gate within  a closed $N$-qubit system, satisfying the controllability conditions \cite{2,3'}. 

\textcolor{black}{The proposed model constitutes the rigorous mathematical formulation of the optimal quantum gate implementation problem  in terms of the generalized Bloch vectors. In this formalism, developed in a general setting in \cite{26}, quantum states, quantum observables and evolution of a $d$-dimensional quantum system are described via vectors, \emph{the Bloch vectors}, in spaces $\mathbb{R}^{d^{2}-1}$ and  $\mathbb{C}^{d^{2}-1}$, respectively. The application of this formalism allows us to formulate the optimal problem for the implementation of an arbitrary $N\geq1$-qubit quantum gate in the unified form \emph{general} within the mathematical theory of optimal control \cite{6,6',7,9,10}. } 

\textcolor{black}{In contrast to the optimal models in \cite{16,18,21,19}, the cost functional of the proposed model has both the terminal part, describing explicitly the  reduction of the  Bloch vector $u(t,t_{0})\in\mathbb{C}^{d^{2}-1}, \ d=2^{N},$ of the unitary evolution operator of an $N$-qubit quantum system at the final time to the Bloch vector $u_{\text{gate}}\in\mathbb{C}^{d^{2}-1}, \ d=2^{N},$  of an $N$-qubit gate to be generated, and also the integral part, specifying the energy costs for gate implementation.}

\textcolor{black}{By extending the Pontryagin principle \cite{6,6'} to the case of complex variables\footnote{See also \cite{5', 13"}.} and using the special scaling of the integral part of the cost functional, \textcolor{black}{we further derive analytically the necessary optimality conditions within the proposed model and} reduce the optimal control synthesis to the solution of the two-point boundary value problem (BVP) for the system of ordinary differential equations (ODEs). The latter can be explored numerically via any of the known BVP solvers for ODEs.}

Several numerical experiments, conducted for the implementation within the developed optimal model of the mostly used $ N = 1, \, 2, \, 3 $ qubit gates\footnote{Within quantum circuits, high-level gates ($N>3$) are sometimes \cite{25'} approximated by a sequence of lower-level gates.}, demonstrate  that this new model leads to the implementation of quantum gates with a high degree of precision. \textcolor{black}{Note that, within the theoretical optimal models proposed in \cite{14, 16, 18, 21, 22, 19, 19'}, the implementation of three-qubit gates has not been considered.}

\textcolor{black}{We stress that the developed optimal model -- as it is formulated below by equations in (\ref{BFCS_model_statement}) --   has the mathematical form, which is standard for the mathematical control theory; can be used for the implementation of an arbitrary\footnote{In \cite{14,16,18,21,22,19}, the implementation of only one-qubit and two-qubit gates was analyzed.} $N=1,2,3,...$ qubit  gate  and can be analyzed numerically by most of the general numerical control-search methods and their combinations used in \cite{14,16,18,21,22,19}. Namely, by: (i) the Krotov method used in \cite{14}; (ii) the gradient projection method used in \cite{16}; (iii) the dual annealing algorithm (DAA) used in \cite{18,19}; (iv) GRAPE used in \cite{21,19}; (v) automatic-differentiation based gradient optimization used in \cite{22}.}

\textcolor{black}{However, compared with the approaches used in \cite{14,18,21,22,19}, the approach, developed in the present article, sharply separates the analytical stage (by finding the necessary optimality conditions) and the numerical stage of the optimal problem analysis. Namely, due to the generalized Bloch vectors formalism, the extension of the Pontryagin principle to the case of complex variables and the use of a special scaling in the cost functional, the developed optimal gate implementation control problem (\ref{BFCS_model_statement}) is first reduced
analytically to the two-point boundary-value ODEs problem (\ref{BFCS_final_equations}) having the unified form for any $N$-qubit gate, and only after that it is analyzed numerically via a standard ODEs BVP solver. This avoids a prior piecewise-constant or "bang-bang" parametrization of the controls.}

The article is organized as follows. 

In Section \ref{BFCS_section_preliminaries} we shortly outline the main issues of the generalized Bloch vectors formalism \cite{26} for the description of quantum states, quantum observables and evolution in time of a finite-dimensional quantum system. 

In Section \ref{BFCS_section_general_optimal_gates_realization}, based on the generalized Bloch vectors formalism, we develop a new model for the optimal generation of an arbitrary $N$-qubit gate within any $N$-qubit closed quantum system satisfying the controllability conditions. The necessary conditions for optimality of the solution of the developed optimal model are formulated via the extension of the Pontryagin principle \cite{6,10} to the complex variables case.

In Section \ref{BFCS_section_numerical}, we present the numerical solutions of the developed optimal gates for a variety of quantum gates. The investigated cases include: (i) the one-qubit gates NOT, the Hadamard gate H, the phase gate S, the $\frac{\pi}{8}$ gate T; (ii) the two-qubit gates CNOT, C-Z; (iii) the three-qubit Toffoli gate. 
The derived numerical results indicate that the developed optimal model leads to the implementation of $N$-qubit gates with a high degree of precision. 

In Section \ref{BFCS_section_conclusion}, we summarize the main results of the present article.

\section{Preliminaries: the generalized Bloch vectors formalism}\label{BFCS_section_preliminaries}
For our further consideration, let us shortly introduce the main issues of the generalized Bloch vectors formalism \cite{26} for the description of quantum observables, quantum states and evolution of a finite-dimensional quantum system. \textcolor{black}{Within this general formalism, the generalized Gell-Mann representation of a $d\geq3$ dimensional quantum state, used in \cite{27,28, 24,25}, corresponds to the representation of a particular type.} 

\subsection{Generalized Bloch vectors}
Let $ \mathcal{L}_{\mathcal{H}_d} $ be the vector space of all linear operators, say $ X $, on the complex Hilbert space  $ \mathcal{H}_{d} $ of a finite dimension $ d\geq2 $, equipped with the scalar product
\begin{equation}
    \langle X_{i},X_{j}\rangle_{\mathcal{L}_{\mathcal{H}_{d}}}:=\mathrm{tr}\left[  X_{i}^{\dagger}X_{j}\right]. \label{BFCS_scalar_product}
\end{equation}
Denote by 
\begin{equation}\label{BFCS_general_basis}
    \begin{split}
        \mathfrak{B}_{\Upsilon_{d}}  &:= \left\{\mathbb{I}_{d}, \, \Upsilon_{d}^{(k)}\in\mathcal{L}_{\mathcal{H}_d}| \, k = 1, \, \ldots, \, d^{2}-1\right\} \\
        \Upsilon_{d}^{(k)}  &= \left(\Upsilon_{d}^{(k)}\right)^{\dagger} \neq 0, \quad \mathrm{tr}\left[  \Upsilon_{d}^{(k)}\right] = 0, \quad \mathrm{tr}
        \left[\Upsilon_{d}^{(k)}\Upsilon_{d}^{(m)}\right] = 2 \delta_{km},
    \end{split}
\end{equation}
the basis in $ \mathcal{L}_{\mathcal{H}_{d}} $, consisting of the identity operator 
$\mathbb{I}_{d}$ on $\mathcal{H}_{d}$ and the tuple 
\begin{equation}\label{BFCS_basis_tuple}
    \Upsilon_{d} := \left(\Upsilon_{d}^{(1)}, \, \ldots, \, \Upsilon_{d}^{(d^{2}-1)}\right)
\end{equation}
of Hermitian traceless operators mutually orthogonal in $ \mathcal{L}_{\mathcal{H}_{d}} $.

For any operator $ X \in \mathcal{L}_{\mathcal{H}_{d}} $
the decomposition over the basis $\mathfrak{B}_{\Upsilon_{d}}$ is given by 
\begin{equation}\label{BFCS_X_decomposition_over_basis}
    \begin{split}
        &X = x^{(0)} \mathbb{I}_d + \sqrt{\frac{d}{2}} \sum\limits_{j=1}^{d^{2}-1} x_{\Upsilon_{d}}^{(j)} \Upsilon_{d}^{(j)}, \quad x^{(0)}=\frac{\mathrm{tr}\left[X\right]}{d} \in \mathbb{C}, \quad x_{\Upsilon_{d}}^{(k)}  =\frac{1}{\sqrt{2d}}\mathrm{tr} \left[\Upsilon_{d}
^{(k)}X \right] \in\mathbb{C}, \\
        &x_{\Upsilon_{d}} := \left(x_{\Upsilon
_{d}}^{(1)}, \, \ldots, \, x_{\Upsilon_{d}}^{(d^{2}-1)}\right)^{T} \in\mathbb{C}^{d^{2}-1},
    \end{split} 
\end{equation}
where the vector $ x_{\Upsilon_{d}} 
\in\mathbb{C}^{d^{2}-1} $ in representation \eqref{BFCS_X_decomposition_over_basis} satisfies the relation
\begin{equation}\label{BFCS_norm_vec_decomposition}
    \frac{1}{d}\mathrm{tr}\left[X^{\dagger}X\right] = \frac{1}{d} \left\Vert X\right\Vert_{\mathcal{L}_{d}}^{2} = |x^{(0)}|^2 + \left\Vert x_{\Upsilon_{d}}\right\Vert_{\mathbb{C}^{d^{2}-1}}^{2}
\end{equation}
and is referred to \cite{26} as the generalized Bloch vector of an operator $X\in\mathcal{L}_{\mathcal{H}_{d}}$.
For any Hermitian operator $ H \in \mathcal{L}_{\mathcal{H}_{d}} $,  its Bloch vector in decomposition \eqref{BFCS_X_decomposition_over_basis} is real-valued $ h_{\Upsilon_{d}} \in \mathbb{R}^{d^{2}-1} $.

In case of a unitary operator $ U\in\mathcal{L}_{\mathcal{H}_{d}} $, the decomposition \eqref{BFCS_X_decomposition_over_basis} takes the form 
\begin{equation}\label{BFCS_U_decomposition_over_basis}
    U=u^{(0)}\mathbb{I}_d + \sqrt{\frac{d}{2}} \sum\limits_{j=1}^{d^{2}-1} u_{\Upsilon_{d}}^{(j)} \Upsilon_{d}^{(j)},
\end{equation}
where complex components $ u^{(0)} $ and $ u_{\Upsilon_d}^{(j)} $ satisfy the relations 
\begin{equation}\label{BFCS_first_integrals}
    \begin{split}
        & |u^{(0)}|^{2}+\left\Vert u_{\Upsilon_d}\right\Vert_{\mathbb{C}^{d^{2}-1}}^{2} = 1, \\[0.1cm]
        & u^{(0)}\overline{u}_{\Upsilon_d}^{(j)}+\overline{u}^{(0)}u_{\Upsilon_d}^{(j)} +\sqrt{\frac{d}{2}}\sum_{k,m}\left( g_{kmj}^{(\Upsilon_{d})}+if_{kmj}^{(\Upsilon_{d})}\right)  u_{\Upsilon_d}^{(k)}\overline{u}_{\Upsilon_d}^{(m)}=0, \quad j=1, \, \ldots, \, (d^{2}-1),
    \end{split}
\end{equation}
with the real-valued constants  
\begin{equation}\label{BFCS_structural_constants}
    g_{kml}^{(\Upsilon_{d})} = \frac{1}{4}\mathrm{tr}\left\{  \left(
    \Upsilon_{d}^{(k)}\circ \Upsilon_{d}^{(m)}\right)  \Upsilon_{d}^{(l)}\right\},\text{ \ \ }f_{kml}^{(\Upsilon_{d})}=\frac{1}{4i}\mathrm{tr} \left\{\left[\Upsilon_{d}^{(k)},\Upsilon_{d}^{(m)}\right]  \text{ }\Upsilon_{d}^{(l)}\right\},
\end{equation}
which are, correspondingly,  symmetric and antisymmetric under the permutation of indices. Here, $A\circ B:=AB+BA,\text{ }\forall A,B\in \mathcal{L}_{\mathcal{H}_d} $.

Among the operator bases of type \eqref{BFCS_general_basis} one of the most used and applied is the basis where the tuple \eqref{BFCS_basis_tuple} consists of the generalized Gell-Mann operators \cite{27,28,24,25}, constituting the high-dimensional extensions of the Pauli operators 
$(\sigma_{1}, \, \sigma_{2}, \, \sigma_{3}) $ on $\mathbb{C}^{2}$ and the Gell-Mann operators on~$\mathbb{C}^{3}$ and having the form:
\begin{equation}\label{BFCS_Gellman_basis}
\Lambda_{d}=(\Lambda_{12}^{(\text{sym})}, \, \ldots, \, \Lambda_{1d}^{(\text{sym})}, \, \ldots, \, \Lambda_{(d-1)d}^{\text{(sym)}}, \, \Lambda_{12}^{(\text{asym})}, \, \ldots, \Lambda_{1d}^{(\text{asym})}, \, \ldots, \, \Lambda_{d-1, d}^{(\text{asym})}, \, \Lambda_{1}^{(\text{diag})}, \, \ldots, \, \Lambda_{d-1}^{(\text{diag})}),
\end{equation}
where 
\begin{equation}\label{BFCS_Gellman_basis2}
    \begin{split}
        & \Lambda_{mk}^{(\text{sym})} = \left\vert m \right\rangle \left\langle k\right\vert
        +\left\vert k\right\rangle \left\langle m\right\vert, \quad 1\leq m<k\leq d, \\[0.1cm]
        & \Lambda_{mk}^{(\text{asym})} = -i \left\vert m\right\rangle \left\langle
        k\right\vert + i \left\vert k\right\rangle \left\langle m\right\vert, \quad 1\leq m<k\leq
        d, \\[0.1cm]
        & \Lambda_{l}^{(\text{diag})} = \sqrt{\frac{2}{l(l+1)}}\left(  \sum\limits_{j = 1}^{l}
        \left\vert j\right\rangle \left\langle j\right\vert -l\text{ }\left\vert
        l+1\right\rangle \left\langle l+1\right\vert \right)  ,\text{ \ \ }1\leq l\leq
        d-1,
    \end{split}
\end{equation}
and $\{\left\vert j\right\rangle \in\mathbb{C}^{d}\}_{j = 1}^{d}$ is the computational basis in $ \mathbb{C}^{d} $. 

\subsection{Unitary evolution}
Consider now the formalism introduced in Section 2.1 for the description of the unitary evolution of a closed quantum system. Let
\begin{equation}
    H_{d}(t): \mathcal{H}_{d} \rightarrow \ \mathcal{H}_{d}, \quad H_{d}(t) = H_{d}^{\dagger}(t),
\end{equation}
be the Hamiltonian  of a $d$-dimensional quantum system, depending, in general, on time. The unitary time evolution of a quantum system state $\rho_{d}(t)$ is specified by the relation
\begin{equation}\label{BFCS_evolution_relation}
    \rho_{d}(t) = U_{d}(t, t_0) \, \rho_{d}(t_0) \, U_{d}^{\dagger}(t, t_0), \quad t\geqslant t_0,
\end{equation}
where the unitary operator $ U_{d}(t, t_0):\mathcal{H}_{d} \rightarrow \ \mathcal{H}_{d} $ constitutes the solution of the Cauchy problem 
\begin{equation}\label{BFCS_Cauchy_problem}
    i\frac{d}{dt}U_d(t,t_{0}) = H_{d}(t) U_d(t,t_{0}), \quad t > t_{0}, \quad  U_d(t_{0},t_{0})=\mathbb{I}_{d}.
\end{equation}
Using decompositions \eqref{BFCS_X_decomposition_over_basis} and \eqref{BFCS_U_decomposition_over_basis} over a basis \eqref{BFCS_general_basis}:
\begin{equation}
    \begin{split}
        & H_{d}(t) = h^{(0)}(t) \, \mathbb{I}_d + \sqrt{\frac{d}{2}} \sum\limits_{j=1}^{d^{2}-1} h_{\Upsilon_{d}}^{(j)}(t) \Upsilon_{d}^{(j)}, \quad h^{(0)} \in \mathbb{R}, \ h_{\Upsilon_{d}} \in \mathbb{R}^{d^{2}-1}, \\[0.1cm]
        & U_{d}(t,t_0) = u^{(0)}(t,t_{0}) \, \mathbb{I}_{d}+\sqrt{\frac{d}{2}} \sum\limits_{j=1}^{d^{2}-1} u_{\Upsilon_{d}}^{(j)}(t, t_{0}) \Upsilon_{d}^{(j)}, \quad u^{(0)} \in \mathbb{C}, \ u_{\Upsilon_{d}} \in \mathbb{C}^{d^{2}-1}\label{16},
    \end{split}
\end{equation}
and substituting them into  \eqref{BFCS_Cauchy_problem}, we come to the following Cauchy problem for component $ u_0(t,t_0) $ and components  $ u_{\Upsilon_d}^{(j)}(t,t_{0}) $ of the Bloch vector of the unitary evolution operator $ U(t,t_0) $ \cite{26,24}:
\begin{equation}\label{BFCS_dynamics_equations}
    \begin{split}
        & i\frac{d}{dt}u^{(0)}(t,t_{0}) = h^{(0)}(t) \, u^{(0)}(t,t_{0}) + h_{\Upsilon_d}(t)\cdot u_{\Upsilon_d}(t,t_{0}), \\
        & i\frac{d}{ dt}u^{(j)}_{\Upsilon_d}(t,t_{0}) = h^{(0)}(t) \, u^{(j)}_{\Upsilon_d}(t, t_{0}) + u^{(0)}(t,t_{0}) \, h_{\Upsilon_d}^{(j)}(t) +\sqrt{\frac{d}{2}} \sum_{m,k = 1}^{d^{2}-1}\left( g_{kmj}^{(\Upsilon_{d})}+\text{ }if_{kmj}^{(\Upsilon_{d})} \right)
        h^{(k)}_{\Upsilon_{d}}(t) \, u^{(m)}_{\Upsilon_{d}}(t,t_{0}), \\
        & u^{(0)}(t_{0},t_{0}) = 1, \quad  u^{(j)}_{\Upsilon_{d}}(t_{0},t_{0})=0,
    \end{split}
\end{equation}
where $ j=1, \, \ldots, \, (d^{2}-1) $ and the real-valued constants $ g_{kmj}^{(\Upsilon_{d})}, \, f_{kmj}^{(\Upsilon_{d})} $ are defined in \eqref{BFCS_structural_constants}. Here and further, $a \cdot b := \sum\limits_{j}a^{(j)} b^{(j)} $ for both real-valued and complex-valued vectors. 

For any $t>t_{0}$, relations \eqref{BFCS_first_integrals} with $ u^{(0)} = u^{(0)} (t, t_{0}) $ and $ u^{(j)}_{\Upsilon_{d}} = u^{(j)}_{\Upsilon_{d}} (t, t_{0}) $ constitute the first integrals of the system of linear ordinary differential equations \eqref{BFCS_dynamics_equations}.

\section{Optimal implementation of \texorpdfstring{$N$}{N}-qubit gates}\label{BFCS_section_general_optimal_gates_realization}
In this section, we use the formalism presented in Section 2 for the formulation of the optimal gate implementation problem in terms of the Bloch vectors for the unitary evolution operator and system Hamiltonians, free (drift) and control.

Consider the unitary evolution of an $ N $-qubit quantum system with Hamiltonian  $ H_d(t) $, which includes the action on this system of a control. For the $ N $-qubit quantum system, dimension $ d = 2^{N} $. 

In what follows, we omit, for short, indices "$d$"\space and "$\Upsilon_d$"\space at the generalized Bloch vectors of the unitary evolution operator of an $N$-qubit system and system Hamiltonian, also, at constants in \eqref{BFCS_structural_constants}, and will not indicate in Eqs. \eqref{BFCS_dynamics_equations} the dependence of $ u^{(0)}(t,t_{0})$ and $u(t,t_{0}) $ on the initial time moment, which we further take $t_{0} = 0 $.

For an $ N $-qubit quantum system under control, the Hamiltonian has, in general, the form
\begin{equation}\label{BFCS_general_Hamiltonian}
    H(t) = H_{\text{free}} + H_{\text{ctr}}(t),
\end{equation}
where $ H_{\text{free}} $ is  the (drift) Hamiltonian of a $ N $-qubit system without control and $ H_{\text{ctr}} $ is the Hamiltonian, describing the action on this system of a control\footnote{The below index "ctr"\space means the abbreviation of control}. For convenience, we designate the first one as "free", though it may also contain components describing the uncontrollable interaction of the quantum system components too.

We consider 
\begin{equation}\label{BFCS_control_Hamiltonian}
    H_{\text{ctr}}(t) = \sum\limits_{l = 1}^{s} \nu_{\text{ctr}}^{(l)}(t) \, H_{l},
\end{equation}
where $ \left\{ H_{l}\right\}_{l =1}^{s} $ is a set of $ s \leqslant d^{2} $ linearly independent time constant Hamiltonians that describe interactions used to control a quantum system, while $ \nu_{\text{ctr}}^{(j)}(t) $ are real-valued scalar controls we manage. Henceforth, we denote the vector of these controls as 
\begin{equation}
    \nu_{\text{ctr}}(t):=\big(\nu^{(1)}_{\text{ctr}}(t), \, \ldots, \, \nu^{(s)}_{\text{ctr}}(t)\big) \in \mathbb{R}^{s}.\label{20}
\end{equation}

For Hamiltonian \eqref{BFCS_general_Hamiltonian}, the Bloch vector $h(t)$ in decomposition (\ref{16}) admits the representation 
\begin{equation}\label{BFCS_starter}
    h^{(0)}(t) = h^{(0)}_{\text{free}} + h^{(0)}_{\text{ctr}}(t) , \quad h(t) = h_{\text{free}} + h_{\text{ctr}}(t),
\end{equation}
where
\begin{equation}
    h^{(0)}_{\text{free}} = \frac{1}{d}\mathrm{tr}[H_{\text{free}}] \in\mathbb{R}, \quad  h_{\text{free}} = \frac{1}{\sqrt{2d}}\mathrm{tr}[H_{\text{free}} \Upsilon_{d}] \in \mathbb{R}^{d^2-1},
\end{equation}
and
\begin{equation}\label{BFCS_finisher}
    \begin{split}
        & h^{(0)}_{\text{ctr}}(t) = \sum\limits_{l = 1}^{s}  \nu_{\text{ctr}}^{(l)}(t) \, h^{(0)}_{l}, \quad h_{\text{ctr}}(t) = \sum\limits_{l = 1}^{s}  \nu_{\text{ctr}}^{(l)}(t) \, h_{l},  \\
        & h^{(0)}_{l} = \frac{1}{d}\mathrm{tr}[H_{l}] \in \mathbb{R}, \quad  h_{l} = \frac{1}{\sqrt{2d}}\mathrm{tr}[H_{l} \Upsilon_{d}] \in \mathbb{R}^{d^2-1}.
    \end{split}
\end{equation}
For the optimal implementation of an $ N $-qubit quantum gate $ U_{\text{gate}} $, we choose the cost functional of the following general form
\begin{equation}\label{BFCS_general_functional}
    \begin{split}
        &J(U, \nu_{\operatorname{ctr}}) = \frac{1}{2d} \, || U(T) -U_{\text{gate}} ||^2_{\mathcal{L}_{\mathcal{H}_d}} + \frac{\varepsilon}{2}\int\limits_{0}^{T} \sum\limits_{l = 1}^{s} w_{l} \left(\nu_{\text{ctr}}^{(l)}(t)\right)^{2} \, dt,
    \end{split}
\end{equation}
where $ T $ is the duration of the physical implementation of an $ N $-qubit gate $  U_{\text{gate}} $, 
\begin{equation}
    w_{l}: =\big(h^{(0)}_{l}\big)^{2}+||h_{l}||^2_{\mathbb{R}^{d^2-1}}=\frac{1}{d}||H_{l}||^{2}_{\mathcal{L}_{\mathcal{H}_{d}}} \label{25},
\end{equation}
and ${\varepsilon}$ is a small scaling parameter such that 
\begin{equation}\label{BFCS_integral_elem}
\frac{\varepsilon}{2}\int\limits_{0}^{T} \sum\limits_{l = 1}^{s} w_{l} \left(\nu_{\text{ctr}}^{(l)}(t) \right)^{2} dt \ll 1.
\end{equation}

The terminal part of the cost functional \eqref{BFCS_general_functional} is defined via the distance in space $\mathcal{L}_{\mathcal{H}_{d}}$ and determines the closeness of the unitary operator $U (T)$  at the final time moment $T$ to the target unitary operator $U_{\text{gate}}$, describing the designed gate. Meanwhile, the integral part of the functional \eqref{BFCS_general_functional} describes the energy costs for the gate implementation with respect to the scale~$ \varepsilon $ and relative weights~$ w_{l} $.

Taking into account decomposition \eqref{BFCS_X_decomposition_over_basis} for a designed $ U_{\text{gate}}$ and equalities \eqref{BFCS_norm_vec_decomposition}, we rewrite the terminal part of the cost functional \eqref{BFCS_general_functional} in the form
\begin{equation}\label{BFCS_terminal_functional}
     J_{\operatorname{term}}(u, \nu_{\operatorname{ctr}}) = \frac{1}{2}\left( \left\vert u^{(0)}(T)-u^{(0)}_{\text{gate}} \right\vert ^{2}+\left\Vert u(T)-u_{\text{gate}}\right\Vert _{\mathbb{C}^{d^{2}-1}}^{2}\right),
\end{equation}
where
\begin{equation}
    u_{\text{gate}}^{(0)}=\frac{1}{d}\mathrm{tr}[U_{\text{gate}}] \in\mathbb{C}, \quad  u_{\text{gate}}=\frac{1}{\sqrt{2d}} \mathrm{tr}[U_{\text{gate}} \Upsilon_{d}]\in\mathbb{C}^{d^{2}-1}.
\end{equation}

Note that the terminal functional in \eqref{BFCS_general_functional} does not attain its global zero minimum if $ U(T) = e^{i \alpha(T)} \, U_{\text{gate}} $, $ \alpha(T) \neq 2 \pi k$. Therefore, this functional  provides the optimal generation of a designed  quantum gate $U_{\text{gate}}$ with taking into account its precise global phase. In view of (\ref{BFCS_evolution_relation}), a choice for a designed gate $U_{\text{gate}}$ of some specific global phase is not critical for the unitary evolution of a state $\rho(t)$, but may affect a time $ T $ of generation of this gate and be used, accordingly, for its regulation. 

Based on the Cauchy problem \eqref{BFCS_dynamics_equations} and the cost functional \eqref{BFCS_general_functional}, where the terminal  part reduces to \eqref{BFCS_terminal_functional}, we formulate the following model for the optimal realization during a finite time $T$ of an arbitrary $N$-qubit quantum gate $ U_{\text{gate}} $
\begin{equation}\label{BFCS_model_statement}
\begin{split}
    & i\frac{d}{dt}u^{(0)}(t) = h^{(0)}(t) \,u^{(0)}(t) +h(t)\cdot u(t), \\
    & i\frac{d}{dt}u^{(j)}(t) = h^{(0)}(t) \,u^{(j)}(t)+u^{(0)}(t) \, h^{(j)}(t) +\sqrt{\frac{d}{2}} \sum_{m, \,k = 1}^{d^{2}-1}\left( g_{kmj} + if_{kmj} \right) h^{(k)}(t) \, u^{(m)}(t), \\
    & u^{(0)}(0) = 1, \quad u^{(j)}(0)=0, \quad j=1,...,(d^{2}-1), \\
    & J =  \frac{1}{2}\left( \left\vert u^{(0)}(T)-u^{(0)}_{\text{gate}} \right\vert ^{2}+\left\Vert u(T)-u_{\text{gate}}\right\Vert_{\mathbb{C}^{d^{2}-1}}^{2}\right) + \frac{\varepsilon}{2}\int_{0}^{T} \sum\limits_{l = 1}^{s} \left(\big(h^{(0)}_{l}\big)^{2}+||h_{l}||^2_{\mathbb{R}^{d^2-1}}\right) \left(\nu_{\text{ctr}}^{(l)}(t)\right)^{2} \, dt  \rightarrow \text{min},
\end{split}
\end{equation}
where $ d=2^{N} $, the small parameter $\varepsilon$ satisfies  relation \eqref{BFCS_integral_elem} and the constants $ g_{kmj}, \, f_{kmj}$ are defined in \eqref{BFCS_structural_constants}.

This optimal model has a unified form general for the mathematical control theory and can be applied for the implementation of an arbitrary $ N \geq 1 $ qubit gate within a unitary evolution of a closed $N$-qubit system {\color{black} which is controllable. The latter means that, under a suitable choice of controls and a time interval, a needed result of the unitary evolution of this system can be attained.

Recall that a closed finite-dimensional quantum system is controllable
in the above sense if and only if the dynamical Lie algebra 
\begin{equation}\label{dynamical_algebra}
\begin{split}
    &\mathcal{L}(iH_0,iH_1,\ldots,iH_s)
:=
\operatorname{span}
\left(
iH_j,\;
[iH_j,iH_k],\;
[iH_j,[iH_k,iH_\ell]],\;\ldots
\right), \\
& j,k,\ell\in\{0,1,\ldots,s\},
\end{split}
\end{equation}
generated
by operators  
$iH_1,\ldots,iH_s$ in~\eqref{BFCS_control_Hamiltonian} and operator $iH_0 := iH_{\operatorname{free}}$, has dimension $d^2$ if at least one of the Hamiltonians
$ H_{0}, H_1,\ldots,H_s$ has a nonzero trace, and $(d^2-1)$ if
all these Hamiltonians are traceless. This criterion is known as the Lie algebra rank condition (LARC).

We note that the dimension of the above algebra~$ \mathcal{L} $ can be determined numerically via the iterative generation of commutators and the subsequent selection of their maximal linearly independent subset.}

By writing out the system of linear ordinary differential equations in  \eqref{BFCS_model_statement} via the real and imaginary parts of functions $u^{(0)}(t)$, $u(t)$ and applying further the Pontryagin formalism \cite{6, 10}, we come to the following statement.

\begin{theorem}\label{BFCS_theorem_with_conds}
The solution of the optimal model \eqref{BFCS_model_statement} satisfies the following necessary optimality conditions
\begin{equation}\label{BFCS_Pontryagins_equations}
    \begin{split}
        & \frac{d}{dt}u^{(0)}(t) = \frac{\partial \mathrm{H}}{\partial \overline{p}^{(0)}}, \quad \frac{d}{dt} u(t) = \frac{\partial \mathrm{H}}{\partial \overline{p}}, \quad u^{(0)}(0)=1, \quad u(0)=0, \\
        & \frac{d}{dt}p^{(0)}(t) = -\frac{\partial \mathrm{H}}{\partial \overline{u}^{(0)}}, \quad
        \frac{d}{dt}p(t) = -\frac{\partial \mathrm{H}}{\partial \overline{u}}, \quad p^{(0)}(T)=-{u}_{\operatorname{gate}}^{(0)}, \quad p(T)=- {u}_{\operatorname{gate}}, \\[0.2cm]
        & \frac{\partial \mathrm{H}}{\partial \nu_{\operatorname{ctr}}}=0,
    \end{split}
\end{equation}
where $ u^{(0)}, \, p^{(0)} \in\mathbb{C} $; $ u, \, p\in\mathbb{C}^{d^{2}-1} $; $ \nu_{\operatorname{ctr}} \in \mathbb{R}^{s} $ and  
\begin{equation}\label{BFCS_Pontryagins_Hamiltonian}
    \begin{split}
         \mathrm{H} &= \varepsilon \sum\limits_{l}\left(\nu_{\text{ctr}}^{(l)}\, (t)\right)^{2} \left(\big(h^{(0)}_{l}\big)^{2}+||h_{l}||^2_{\mathbb{R}^{d^2-1}}\right) + 2 \, \mathrm{Im} \left(\overline{p}^{(0)}(t) \big[h^{(0)}(t) \, u^{(0)}(t) +h(t)\cdot u(t)\big]\right) +\\ &+ 2 \, \mathrm{Im}\left(\overline{p}(t)\cdot \big[h^{(0)}(t) \, u(t) +u^{(0)}(t) \, h(t)\big]\right) + \sqrt{2 d} \, \mathrm{Im} 
        \Big(\sum_{k, m, j = 1}^{d^{2}-1}\left( g_{kmj}+if_{kmj} \right) \,
        h^{(k)}(t) \, u^{(m)}(t) \, \overline{p}^{(j)}(t) \Big).
    \end{split}
\end{equation}
In Eqs. \eqref{BFCS_Pontryagins_equations}, the derivatives of the Hamiltonian function $\mathrm{H}(\cdot )$ over the complex variables are in the sense of the formal derivative\footnote{That is $\frac{\partial  f(z)}{\partial \overline{z}}:=\frac{1}{2} \Big( \frac{\partial f(z)}
{\partial (\text{Re} \, z)} +i\frac{\partial f(z)}{\partial (\text{Im} \, z)}\Big)$} by Wirtinger \cite{29}. Here, variables $ \text{Re}(p)$ and $ \text{Im}(p) $ are conjugate to $ \text{Re}(u) $ and $ \text{Im}(u) $, respectively.  

\end{theorem}

 Taking into account  \eqref{BFCS_Pontryagins_Hamiltonian}, we represent the necessary optimality conditions  \eqref{BFCS_Pontryagins_equations}, determining the admissible solutions of the optimal model \eqref{BFCS_model_statement} in the form
\begin{equation}
    \begin{cases}\label{BFCS_final_equations}
        &\displaystyle i\frac{d}{dt} u^{(0)}(t) = h^{(0)}(t) \, u^{(0)}(t) + h(t) \cdot u(t), \\
        &\displaystyle i\frac{d}{ dt}u^{(j)}(t)  = h^{(0)}(t) \, u^{(j)}(t) + u^{(0)}(t) \, h^{(j)}(t) +\sqrt{\frac{d}{2}} \sum_{k, m = 1}^{d^{2}-1} \big(g_{kmj}+if_{kmj}\big) \, h^{(k)}(t) \, u^{(m)}(t), \\  
        &\displaystyle i\frac{d}{dt}p^{(0)}(t) = h^{(0)}(t) \, p^{(0)}(t) + h(t) \cdot p(t), \\
        &\displaystyle i\frac{d}{dt}p^{(j)}(t) = h^{(0)}(t) \, p^{(j)}(t) + p^{(0)}(t) \, h^{(j)}(t) + \sqrt{\frac{d}{2}} \sum_{k,m = 1}^{d^{2}-1}\big( g_{kmj} + if_{kmj}\big) \, h^{(k)}(t) \, p^{(m)}(t), \\
        &u^{(0)}(0)=1, \quad u^{(j)}(0)=0, \quad p^{(j)}(T) = -u^{(j)}_{\text{gate}}, \quad p^{(0)}(T)=-u^{(0)}_{\text{gate}}, \\[0.2cm]
        &\varepsilon \, \nu_{\text{ctr}}^{(l)}(t) \left(\big(h^{(0)}_{l}\big)^{2}+||h_{l}||^2_{\mathbb{R}^{d^2-1}}\right)
        + h_{l}^{(0)} \, \text{Im}\Big(\overline{p}^{(0)}(t) \, u^{(0)}(t) +  \overline{p}(t) \cdot u(t) \Big) + \\[0.2cm]
        &\displaystyle +\, h_{l} \cdot \text{Im}\Big(\overline{p}^{(0)}(t) \, u(t) + u^{(0)}(t) \, \overline{p}(t)\Big) + \sqrt{\frac{d}{2}} \, \mathrm{Im} \Big(\sum_{k, m, j = 1}^{d^2-1}\left( g_{kmj}+if_{kmj} \right) \,
        h_{l}^{(k)} \, u^{(m)}(t) \, \overline{p}^{(j)}(t) \Big) = 0,
    \end{cases}
\end{equation}
where $ u^{(0)}(t), u^{(j)}(t), \, \nu_{\text{ctr}}^{(l)}(t)\, \in\mathbb{C} $, $ j=1, \, \ldots, \, (d^{2}-1)$, $ l\,= 1, \,\ldots,\, s $.

 We stress that Theorem \ref{BFCS_theorem_with_conds} provides only necessary conditions for solving the optimal model (\ref{BFCS_model_statement}).
 Taking into account relation \eqref{BFCS_integral_elem}, within our further numerical study in Section \ref{BFCS_section_numerical}, we consider the numerical solution of the boundary value problem (\ref{BFCS_final_equations}) for some small value of parameter $ \varepsilon $ in the cost functional (\ref{BFCS_general_functional}) to be valid for the optimal generation of the corresponding gate if, for this solution, the terminal part of the cost functional in (\ref{BFCS_model_statement}) is close to zero. 

\section{Numerical study}\label{BFCS_section_numerical}
In this Section, for various $N=1,2,3$  qubit gates, we find the numerical solutions of the boundary value problem (\ref{BFCS_final_equations}) with different values $\{0.005,0.05,0.5,5\}$  of the scaling parameter $\varepsilon$ in the cost functional (\ref{BFCS_general_functional}). {\color{black} In what follows, the Lie algebras \eqref{dynamical_algebra}, generated by the Hamiltonians of the considered $ N $-qubit systems, satisfy the Lie algebra rank condition (LARC), discussed after Eq.(\ref{BFCS_model_statement}) in Section~\ref{BFCS_section_general_optimal_gates_realization}. Therefore, these $N$-qubit systems are controllable.} 

In order to develop the unified code valid for all $ N \geq 1 $, for the construction of the Bloch vectors $ u_{\text{gate}} $ and $ h $, standing in \eqref{BFCS_final_equations}, we use the specific operator basis \eqref{BFCS_Gellman_basis}, consisting of the generalized Gell-Mann operators   \eqref{BFCS_Gellman_basis2}. In this case, the constants $g_{kmj},f_{kmj}$,   standing in \eqref{BFCS_final_equations} and defined in general by (\ref{BFCS_structural_constants}), constitute the structure constants of group $\text{SU}(2^{N})$, the values of which for $ N = 1, \,  2, \, 3$, are well-known. 

For finding the numerical solutions of the boundary value problem \eqref{BFCS_final_equations}, we apply the boundary value problem (BVP) solver via using a collocation method with an adaptive mesh provided by the scipy.integrate package function solve\_bvp \cite{30}. We stress that this solver can be replaced by any other suitable numerical tool for the ODEs system~\eqref{BFCS_final_equations}. {\color{black} Since the number of equations in this ODEs system scales as $d^{2}$ and thus grows exponentially with the number $N$ of qubits, the numerical analysis of this system for large $N$ may require more specialized numerical solvers.}

The solver constructs a piecewise polynomial approximation of the solution and applies the Newton method to determine its coefficients, ensuring the equations and boundary conditions are satisfied. If needed, the algorithm refines the mesh by adding points in regions with large errors and iterates until convergence. The first integrals \eqref{BFCS_first_integrals} provide an easy way to verify that the solver has not failed. Due to the last equation in \eqref{BFCS_final_equations}, this system of ordinary differential equations becomes ill-conditioned as $ \varepsilon $ decreases. For the successful convergence, this necessitates a more accurate initial approximation. To address this, we use an iterative approach by obtaining first the solutions for larger $\varepsilon$ values and further using them as the initial approximations for subsequent iterations with smaller values of the parameter $ \varepsilon $. The generation time $T$ is also selected iteratively: the calculations are first performed for a large value $T$, after that the time of the best obtained local minimum of the terminal functional is selected as the new value of $T$ and this process is repeated until a sufficiently effective generation time is found. 

{\color{black} Each problem considered below was solved numerically using different initial guesses a sufficient number of times to assess the sensitivity of the solution to the initial guess choice.  For each initial guess, the value at every mesh point was sampled from the Gaussian distribution $ \mathcal{N}(0,\sigma^{2}) $ with $ \sigma=0.1 $. The resulting solutions showed no significant sensitivity to the choice of initial guess. Below, for each considered $N$-qubit case, we present the result yielding the smallest value of the corresponding terminal functional in Eq.~(\ref{BFCS_model_statement}).}

In what follows in Subsections \ref{BFCS_subsection_1qgates}, \ref{BFCS_subsection_2qgates} and \ref{BFCS_subsection_3qgates}, we present the numerical results on the optimal implementation within model \eqref{BFCS_model_statement} of various one-qubit, two-qubit and three-qubit gates, with all the information about these numerical results given in Figures  \ref{fig:BFCS_one_qubit_pictures}, \ref{fig:BFCS_two_qubit_pictures},  and \ref{fig:BFCS_three_qubit_pictures}, respectively.

In each of the considered cases, we present the plot for the value of the terminal part of the functional for $t\leq T$ -- in order to provide an insight into the dynamics of this value for $ \varepsilon = 5; \, 0.5; \, 0.05; \, 0.005 $, and the plot for the time dependence of optimal controls for $ \varepsilon = 0.005 $. 

For one-qubit gates, we also show in Figure \ref{fig:BFCS_one_qubit_pictures}, the evolution in time of the non-zero components of the Bloch vectors $ u(t) $ for $ \varepsilon = 0.005 $.    However,  for $ N > 1 $, similar plots are not provided in view of the large number  of the Bloch vector components. 

Where the curves are close to each other, some intermediate values are omitted. 
For implementation of the considered one-qubit, two-qubit and three-qubit gates, the values of the terminal functional at time $T$ are specified in Tables \ref{tab:BFCS_1q}, \ref{tab:BFCS_2q} and  \ref{tab:BFCS_3q}, respectively. 

\textcolor{black}{The numerical results presented below clearly indicate that, for small values of the scaling parameter $ \varepsilon$ in the cost functional (\ref{BFCS_general_functional}),  the developed optimal model (\ref{BFCS_model_statement}) enables the implementation of one-qubit, two-qubit and three-qubit gates with a high degree of precision, as evidenced by the small final values of the terminal functional and the corresponding values of the average gate fidelity (AGF), which are close to 1.} 

\textcolor{black}{Recall that the average gate fidelity $ F_{\mathrm{AGF}}(U(T),U_{\mathrm{gate}}) $ between the target unitary $ U_{\mathrm{gate}} $ and implemented $U(T)$ is defined~\cite{Nielsen2002} by  the relation
\begin{equation}
    F_{\mathrm{AGF}}(U(T), U_{\operatorname{gate}})
    := \frac{
    \left|\operatorname{tr}\left[
    U_{\operatorname{gate}}^{\dagger} \, U(T)
    \right]\right|^{2}+d}{d(d+1)} \leq1,\label{agf}
\end{equation}
where in our case $d=2^{N} $. Since the terminal part of the cost functional~\eqref{BFCS_general_functional} can be rewritten as 
\begin{equation}
    J_{\operatorname{term}} = \frac{1}{2 d}\left\|U(T)-U_{\text{gate }}\right\|_{\mathcal{L}_{\mathcal{H}_{d}}}^2=1-\frac{1}{d} \operatorname{Re} \left(\operatorname{tr}\left[U_{\text {gate }}^{\dagger} U(T)\right] \right)
\end{equation}
and $ \operatorname{Re}^{2} \left(\operatorname{tr}\left[U_{\text {gate }}^{\dagger} U(T)\right] \right) \leqslant \left| \operatorname{tr}\left[U_{\text {gate }}^{\dagger} U(T)\right] \right|^{2}$, the average gate fidelity obeys the following lower bound
\begin{equation}\label{BFCS_fidelityinequality}
\begin{aligned}
F_{\mathrm{AGF}}(U(T), U_{\operatorname{gate}})
&\geq F^{(\operatorname{bound})}_{\mathrm{AGF}}
(U(T), U_{\operatorname{gate}}) 
:= \frac{d(1-J_{\mathrm{term}})^2+1}{d+1} \\
&= 1-\frac{d}{d+1}
\left(2J_{\mathrm{term}}-J_{\mathrm{term}}^2\right).
\end{aligned}
\end{equation}
For the numerical runs in the subsections below, we provide both the terminal functional values and the corresponding lower bounds on the average gate fidelity obtained from the right-hand side of inequality~\eqref{BFCS_fidelityinequality}.
}

\subsection{One-qubit gates}\label{BFCS_subsection_1qgates}

As examples for $N=1$, we consider the optimal implementation of the following one-qubit  gates \cite{23} \begin{equation}\label{gates1}
    \begin{split}
        &\text{NOT} = \sigma_{1} = |0\rangle\langle1| + |1\rangle\langle0|, \\[0.1cm]
        &\text{H} = \frac{1}{\sqrt{2}} \big( |0\rangle \langle 0| + |0\rangle \langle 1| + |1\rangle \langle 0| - |1\rangle \langle 1|  \big),\\
        &\text{S} = \sqrt{Z} = |0\rangle \langle 0| + i |1\rangle \langle 1|, \\[0.2cm] 
        &\text{T} = \sqrt{S} = |0\rangle \langle 0| + e^{i \frac{\pi}{4}} |1\rangle \langle 1|,
    \end{split}
\end{equation}
within the unitary evolution of the one-qubit quantum system with the traceless Hamiltonian
\begin{equation}\label{BFCS_one_qubit_hamiltonian}
    H(t) = \frac{\omega}{2} \sigma_{3} + \alpha \, \sigma_{2} +  \nu(t) \, \sigma_{1},
\end{equation}
satisfying the controllability conditions specified in \cite{2} and used for the experimental generation of one-qubit gates in \cite{31} (see there Eq. (13)). 

Here: (i) $ |0\rangle, \, |1\rangle$ are the elements of the computational basis in $\mathbb{C}^{2} $; (ii) $\sigma_{1}, \sigma_{2}, \sigma_{3}$  are the Pauli operators; (iii) parameters $ \omega,\alpha >0$ and function $ \nu(t)\in\mathbb{R}$ is a time dependent scalar control. 

In terms of notations used in \eqref{BFCS_general_Hamiltonian}--\eqref{BFCS_finisher}, Hamiltonian \eqref{BFCS_one_qubit_hamiltonian} consists of
the free Hamiltonian $H_{\text{free}}$ and the control Hamiltonian $H_{\text{ctr}}$ with the following components within the decomposition in the operator basis {$\left\{\mathbb{I}_{d}, \sigma_{1}, \sigma_{2},\sigma_{3} \right\} $:
\begin{equation}\label{BFCS_one_qubit}
    \begin{aligned}
        &H_{\text{free}} = \frac{\omega}{2} \sigma_{3} + \alpha \, \sigma_{2}, \quad &&h_{\text{free}}^{(0)} = 0, \quad &&h_{\text{free}} = \big(0, \, \alpha, \, \frac{\omega}{2} \big); \\[0.2cm]
        &H_{\text{ctr}} = \nu(t) \sigma_{1}, \quad &&h_{\text{ctr}}^{(0)} = 0,  \quad &&h_{\text{ctr}}(t) = \nu(t) \big(1, \, 0, \, 0 \big).
    \end{aligned}
\end{equation}
Therefore, for the one-qubit Hamiltonian \eqref{BFCS_one_qubit_hamiltonian},  the value of $ h^{(0)}(t) \in \mathbb{R} $ and the Bloch vector $ h(t) \in \mathbb{R}^{3} $  are given by
\begin{equation}
    h^{(0)} = h_{\text{free}}^{(0)} + h_{\text{ctr}}^{(0)} = 0, \quad h(t) = h_{\text{free}} + h_{\text{ctr}}(t) = \big(\nu(t), \, \alpha, \, \frac{\omega}{2} \big). 
\end{equation}

Note that component $ h^{(0)} $ is equal to zero for any time $ t $ since Hamiltonian (\ref{BFCS_one_qubit_hamiltonian}) is traceless. This implies that we are solving the gate generation problem within the special unitary group $ \text{SU(2)} $, and the form of the terminal functional in \eqref{BFCS_general_functional} forces us to introduce for each of the designed one-qubit gates $ U_{\text{gate}}$  presented in (\ref{gates1}) the  global phase correction factor $ e^{i \alpha} $ in such a way that $ \text{det}( e^{i \alpha} U_{\text{gate}}) = 1 $. Therefore, we take: (i) $ \alpha = \frac{\pi}{2} $ for gates NOT and H; (ii) $ \alpha = -\frac{\pi}{4} $ for gate S and (iii) $ \alpha = -\frac{\pi}{8}$ for gate T. 

Taking this and Eqs. (\ref{BFCS_one_qubit}) into account, we come to the following expressions for the Bloch vectors of the phase corrected one-qubit gates:
\begin{equation}
    \begin{aligned}
        &u_{e^{i \frac{\pi}{2}}\text{NOT}}^{(0)} = 0, \quad &&u_{e^{i \frac{\pi}{2}}\text{NOT}} = \big(i, \, 0, \, 0\big);\\[0.2cm]
        &u_{e^{i \frac{\pi}{2}}\text{H}}^{(0)} = 0, \quad &&u_{e^{i \frac{\pi}{2}}\text{H}} = \frac{1}{\sqrt{2}
        } \big(i, \, 0, \, i\big); \\[0.2cm]
        &u_{e^{-i \frac{\pi}{4}}\text{S}}^{(0)} = \frac{1}{\sqrt{2}}, &&u_{e^{-i \frac{\pi}{4}}\text{S}} = \frac{1}{\sqrt{2}} \big(0, \, 0, \, -i\big); \\[0.2cm]
        &u_{e^{-i \frac{\pi}{8}}\text{T}}^{(0)} = \cos\left(\frac{\pi}{8}\right), \quad &&u_{e^{-i \frac{\pi}{8}}\text{T}} = \big(0, \, 0, \, -i \sin\left(\frac{\pi}{8}\right)\big)\label{36}.
    \end{aligned}
\end{equation}
These relations, in particular, imply that, for all considered one-qubit gates, $\text{Im} (u^{(0)}_{\text{gate}}(t)=0, \, \text{Re}(u_{\text{gate}}(t)) =0$, for all $t\in(0,T]$,   though it is not the case for higher dimensions.

By substituting (\ref{BFCS_one_qubit}) and (\ref{36}) into the boundary value problem \eqref{BFCS_final_equations}, taking into account that, for SU(2), the structure constant $g_{kmj}=0$ while  the structure constant $f_{kmj}$ constitutes the Levi-Civita symbol\footnote{The value of $f_{kmj}$ is equal to unity if tuple $(k,m,j)$ constitutes an even permutation of $(1,2,3)$ and  minus unity if  $(k,m,j)$ is an odd permutation of $(1,2,3)$ and zero if values of some indices coincide.}, and using further the numerical method described above, we derive for gates (\ref{36}) the numerical results presented in Figure \ref{fig:BFCS_one_qubit_pictures} and Table \ref{tab:BFCS_1q}.  

\begin{figure}[!htbp]
  \centering
  \begin{subfigure}[t]{0.49\textwidth}
    \centering
    \includegraphics[width=\linewidth]{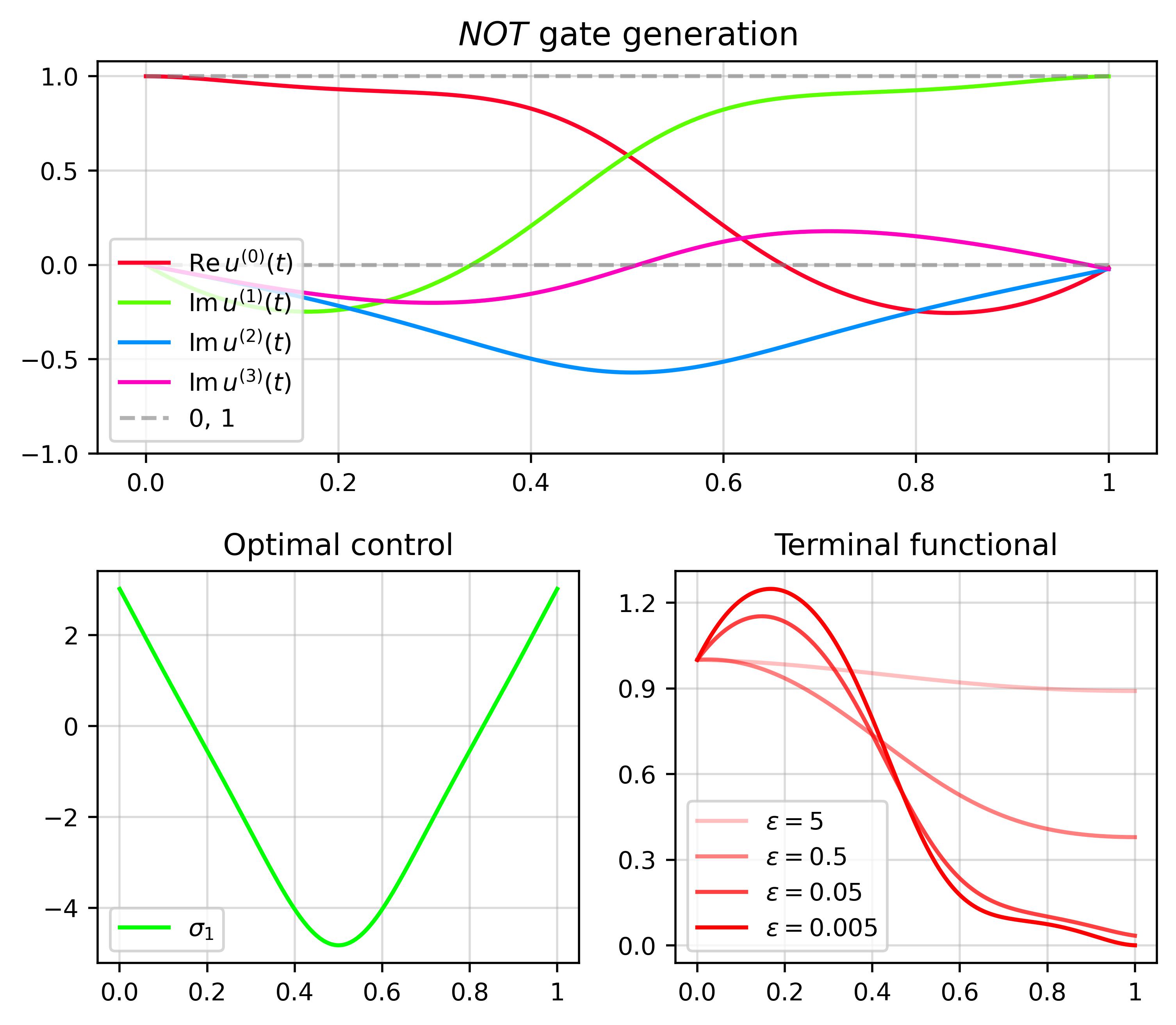}
    \subcaption{}
    \label{fig:BFCS_1a}
  \end{subfigure}
  \hfill
  \begin{subfigure}[t]{0.49\textwidth}
    \centering
    \includegraphics[width=\linewidth]{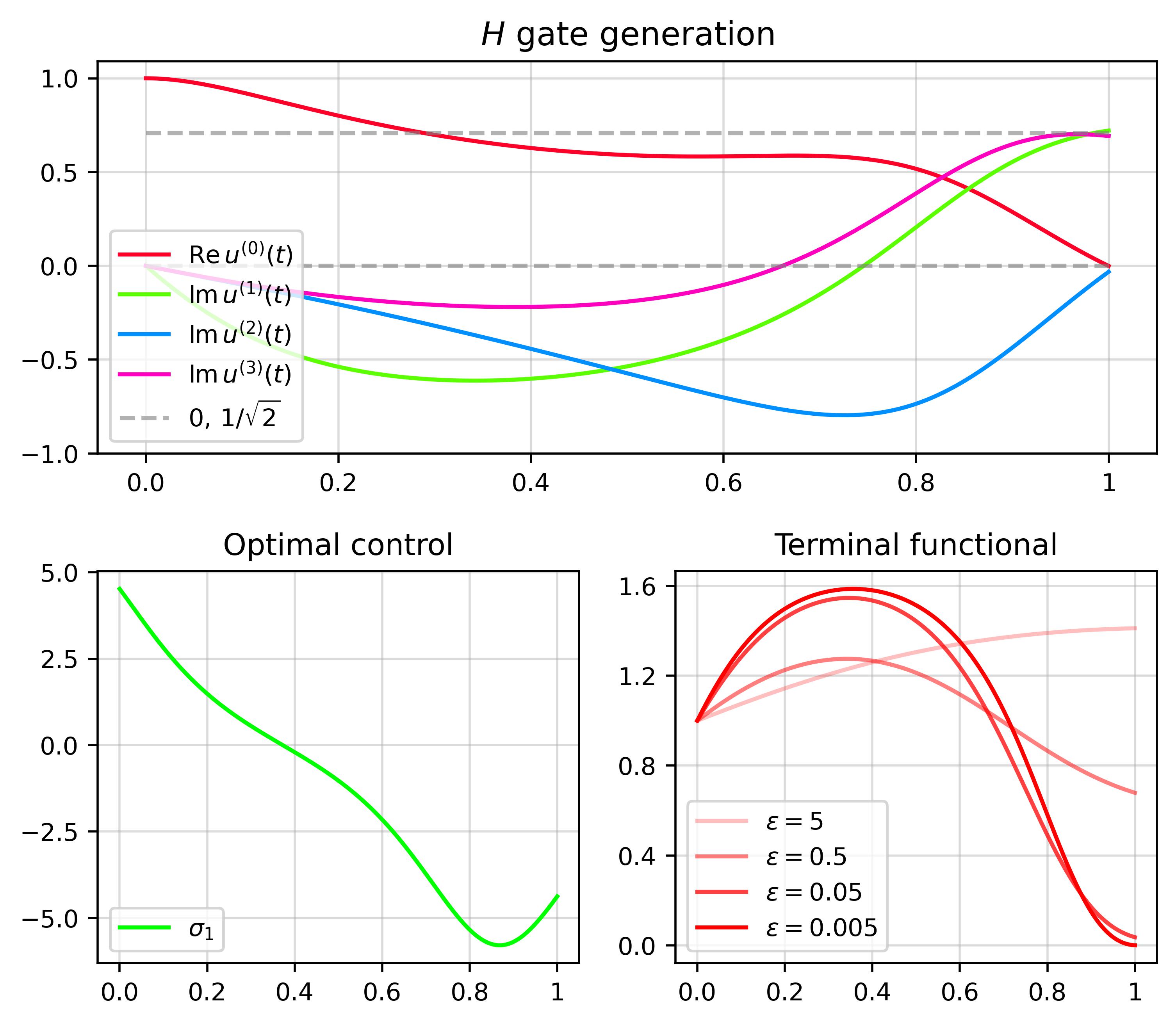}
    \subcaption{}
    \label{fig:BFCS_1b}
  \end{subfigure} \\[0.3cm]
  
  \begin{subfigure}[t]{0.49\textwidth}
    \centering
    \includegraphics[width=\linewidth]{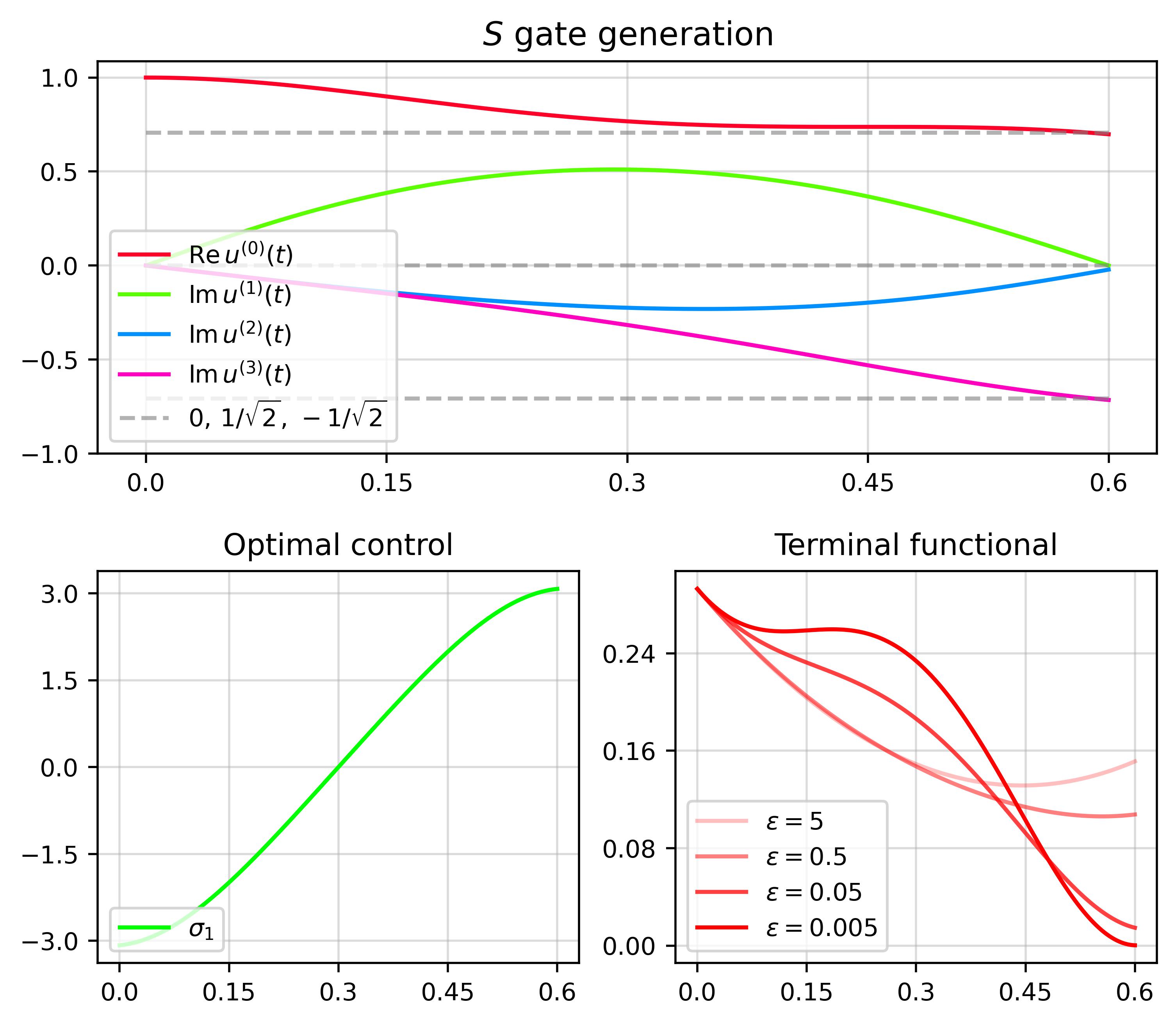}
    \subcaption{}
    \label{fig:BFCS_1c}
  \end{subfigure}
  \hfill
  \begin{subfigure}[t]{0.49\textwidth}
    \centering
    \includegraphics[width=\linewidth]{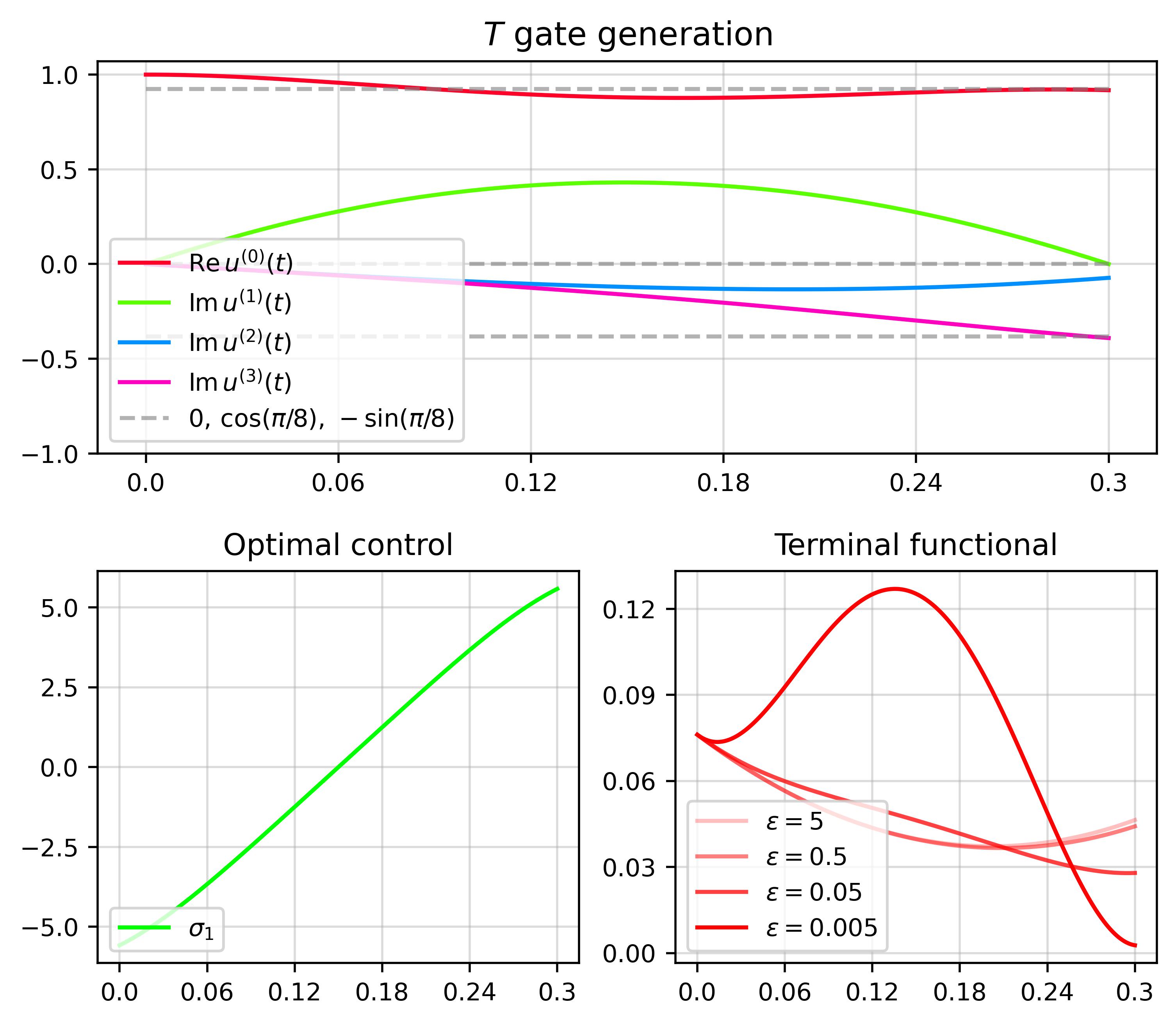}
    \subcaption{}
    \label{fig:BFCS_1d}
  \end{subfigure} \\[0.3cm]
  \caption{Results of the gate generation for (\ref{fig:BFCS_1a}) NOT, (\ref{fig:BFCS_1b}) H, (\ref{fig:BFCS_1c}) S and (\ref{fig:BFCS_1d}) T via system \eqref{BFCS_one_qubit_hamiltonian} with $ \omega = 2 $, $ \alpha = 1 $. Considered gate generation time is $ T = 1 $ for NOT and H gates, $ T = 0.6 $ for S gate and $ T = 0.3 $ for T gate. The plots for the optimal controls and the time dependence of non-zero components of the Bloch vectors $u(t)$ are given for $\varepsilon = 0.005$.} 
  \label{fig:BFCS_one_qubit_pictures}
\end{figure}

\begin{table}[htbp!]
  \centering
  \begin{tabular}{|c|c|c|c|c|c|}
    \hline
    \multirow{2}{*}{\textcolor{black}{\textbf{Gate} $ \boldsymbol{U_{\operatorname{gate}}} $}}
    & \multirow{2}{*}{\textcolor{black}{$\boldsymbol{J_{\mathrm{term}}}$ \textbf{/} $\boldsymbol{F^{(\operatorname{bound})}_{\mathrm{AGF}}}$}}
    & \multicolumn{4}{c|}{\textcolor{black}{\textbf{Scaling parameter} $\boldsymbol{\varepsilon}$}} \\
    \cline{3-6}
    & & $5 \cdot 10^{0}$
      & $5 \cdot 10^{-1}$
      & $5 \cdot 10^{-2}$
      & $5 \cdot 10^{-3}$ \\
    \hline
    \multirow{2}{*}{$e^{i \frac{\pi}{2}}\text{NOT}$}
    & $\textcolor{black}{J_{\mathrm{term}}}$
    & $0.8906$ & $0.3791$ & $0.0342$ & $0.0006$ \\
    \cline{2-6}
    & \textcolor{black}{$F^{(\operatorname{bound})}_{\mathrm{AGF}}$}
    & \textcolor{black}{$0.3413$}
    & \textcolor{black}{$0.5903$}
    & \textcolor{black}{$0.9552$}
    & \textcolor{black}{$0.9992$} \\
    \hline
    \multirow{2}{*}{$e^{i \frac{\pi}{2}}\text{H}$}
    & $\textcolor{black}{J_{\mathrm{term}}}$
    & $1.4110$ & $0.6790$ & $0.0365$ & $0.0007$ \\
    \cline{2-6}
    & \textcolor{black}{$F^{(\operatorname{bound})}_{\mathrm{AGF}}$}
    & \textcolor{black}{$0.4459$}
    & \textcolor{black}{$0.4020$}
    & \textcolor{black}{$0.9522$}
    & \textcolor{black}{$0.9991$} \\
    \hline
    \multirow{2}{*}{$e^{-i \frac{\pi}{4}}\text{S}$}
    & $\textcolor{black}{J_{\mathrm{term}}}$
    & $0.1512$ & $0.1077$ & $0.0147$ & $0.0003$ \\
    \cline{2-6}
    & \textcolor{black}{$F^{(\operatorname{bound})}_{\mathrm{AGF}}$}
    & \textcolor{black}{$0.8136$}
    & \textcolor{black}{$0.8641$}
    & \textcolor{black}{$0.9805$}
    & \textcolor{black}{$0.9996$} \\
    \hline
    \multirow{2}{*}{$e^{-i \frac{\pi}{8}}\text{T}$}
    & $\textcolor{black}{J_{\mathrm{term}}}$
    & $0.0464$ & $0.0442$ & $0.0279$ & $0.0027$ \\
    \cline{2-6}
    & \textcolor{black}{$F^{(\operatorname{bound})}_{\mathrm{AGF}}$}
    & \textcolor{black}{$0.9396$}
    & \textcolor{black}{$0.9424$}
    & \textcolor{black}{$0.9633$}
    & \textcolor{black}{$0.9964$} \\
    \hline
  \end{tabular} 
  \caption{{\color{black} The values of the terminal functional $J_{\mathrm{term}}$ and the corresponding lower bounds $F^{(\operatorname{bound})}_{\mathrm{AGF}}$ on the average gate fidelity within the experiments on one-qubit gates generation with setups described in the caption of Figure~\ref{fig:BFCS_one_qubit_pictures}. Here and throughout the following numerical runs, the BVP solver is used with a tolerance of  $ 10^{-3} $ for both the normalized root-mean-square residuals of the differential equations and the absolute residuals of the boundary conditions. The BVP solver is considered to have converged successfully when the residuals fall below the prescribed tolerances on the adaptive mesh. The initial mesh is uniform; however, the solver may adaptively add nodes in regions where the prescribed accuracy has not been achieved. For all of the provided one-qubit experiments, both the initial and final BVP solver mesh sizes are $500$~nodes.}}
  \label{tab:BFCS_1q}
\end{table}
 
\subsection{Two-qubit gates}\label{BFCS_subsection_2qgates}
In a two-qubit case, we solve numerically the developed optimal model for generating the controlled gates CNOT and CZ:
\begin{equation}
    \begin{split}
        &\text{CNOT} = |0\rangle \langle 0| \otimes \mathbb{I}_{2} + |1\rangle \langle1|\otimes \text{NOT}, \\[0.2cm]
        &\text{CZ} = |0\rangle \langle 0| \otimes \mathbb{I}_{2} + |1\rangle \langle1|\otimes \text{Z},\label{gates2}
    \end{split}
\end{equation}
(where $ Z = \sigma_{3}) $ within the unitary evolution of the two-qubit system, described by the Hamiltonian
generalizing  \eqref{BFCS_one_qubit_hamiltonian} for a two-qubit case:
\begin{equation}\label{BFCS_two_qubit_hamiltonian}
    \begin{split}
        H &= \frac{\omega_{1}}{2} \, \sigma_{3} \otimes \mathbb{I} + \frac{\omega_{2}}{2} \, \mathbb{I} \otimes \sigma_{3} + \alpha \, \mathbb{I} \otimes \sigma_{2} + \beta_{1} \, \sigma_{2} \otimes \sigma_{2} + \beta_{2} \, \sigma_{3} \otimes \sigma_{3} +  \\[0.2cm]
        &+ \nu_{1}(t) \, \sigma_{1} \otimes \mathbb{I} + \nu_{2}(t) \, \sigma_{2} \otimes \mathbb{I} + \nu_{3} (t) \, \mathbb{I} \otimes \sigma_{1}.
    \end{split}
\end{equation}
It retains the free components $\frac{\omega_{i}}{2} \sigma_{3}$ for both qubits and an uncontrolled local component $\alpha \sigma_{2}$ for the second one, while the terms $\sigma_{1} \otimes \mathbb{I}$, $\sigma_{2} \otimes \mathbb{I}$, and $\mathbb{I} \otimes \sigma_{1}$ are manageable. The generation of a non-trivial two-qubit gate requires an interaction between the qubits, therefore, $ H $ contains also $ \beta_{1} \, \sigma_{2} \otimes \sigma_{2} $ and $ \beta_{2} \, \sigma_{3} \otimes \sigma_{3} $. Thus, the stated problem is to manipulate the local controls in such a way that by exploiting this existing interaction to achieve the generation of a required  gate.

For the traceless two-qubit Hamiltonian (\ref{BFCS_two_qubit_hamiltonian}), the component $h^{(0)}=0$,  the free and control Hamiltonians and the non-zero components of their Bloch vectors in $\mathbb{R}^{15}$  are given by \begin{equation}
    \begin{split}
        &H_{\text{free}} =  \frac{\omega_{1}}{2} \, \sigma_{3} \otimes \mathbb{I} + \frac{\omega_{2}}{2} \, \mathbb{I} \otimes \sigma_{3} + \alpha \, \mathbb{I} \otimes \sigma_{2} + \beta_{1} \, \sigma_{2} \otimes \sigma_{2} + \beta_{2} \, \sigma_{3} \otimes \sigma_{3}, \\[0.2cm]
        &h_{\text{free}}^{(2)} = h_{\text{free}}^{(12)} = \frac{\alpha}{\sqrt{2}}, \quad h^{(5)}_{\text{free}} = -h^{(7)}_{\text{free}} = - \frac{\beta_{1}}{\sqrt{2}}, \quad h_{\text{free}}^{(13)} = \frac{2 \beta_{2} + \omega_{2}}{2 \sqrt{2}}, \\[0.2cm]
        &h_{\text{free}}^{(14)} = \frac{2 \beta_{2} + 2 \omega_{1} - \omega_{2}}{2 \sqrt{6}}, \quad h_{\text{free}}^{(15)} = \frac{-2 \beta_{2} + \omega_{1} + \omega_{2}}{2 \sqrt{3}},
    \end{split}
\end{equation}
\begin{equation}
    \begin{split}
        &H_{\text{ctr}}(t) = \nu_{1}(t) \, \sigma_{1} \otimes \mathbb{I} + \nu_{2}(t) \, \sigma_{2} \otimes \mathbb{I} + \nu_{3} (t) \, \mathbb{I} \otimes \sigma_{1}, \\[0.2cm]
        &h_{\text{ctr}}^{(3)}(t) = h_{\text{ctr}}^{(9)}(t) = \frac{1}{\sqrt{2}} \nu_{1}(t), \quad h_{\text{ctr}}^{(4)}(t) = h_{\text{ctr}}^{(10)}(t) = \frac{1}{\sqrt{2}} \nu_{2}(t), \quad h_{\text{ctr}}^{(1)}(t) = h_{\text{ctr}}^{(11)}(t) = \frac{1}{\sqrt{2}} \nu_{3}(t). 
    \end{split}
\end{equation}

\begin{figure}[!htbp]
  \centering
  \begin{subfigure}[t]{\textwidth}
    \centering
    \includegraphics[width=\linewidth]{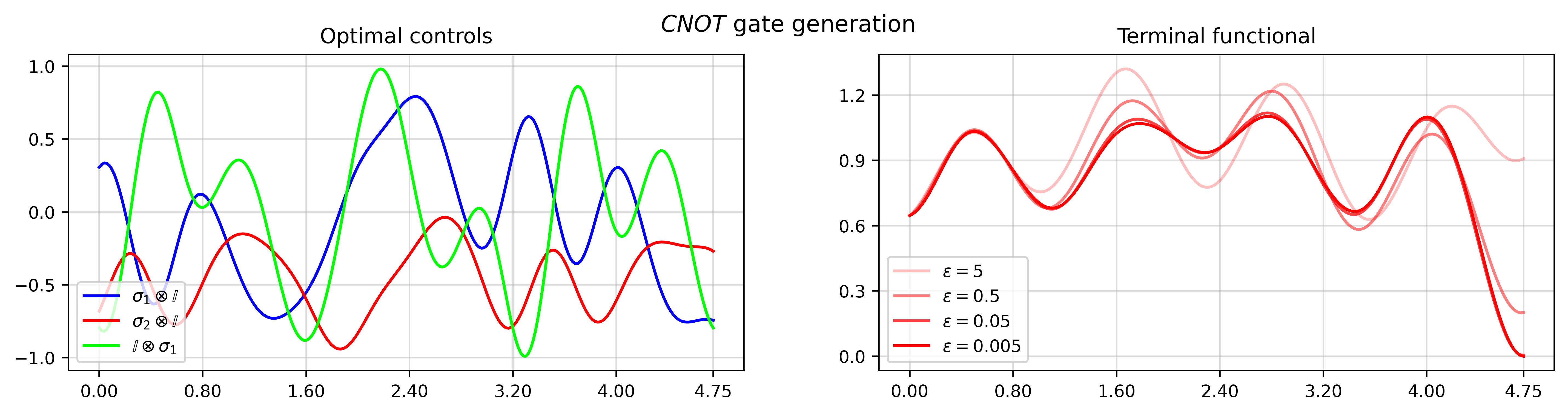}
    \subcaption{}
    \label{fig:BFCS_2a}
  \end{subfigure}\\[0.3cm]
  \begin{subfigure}[t]{\textwidth}
    \centering
    \includegraphics[width=\linewidth]{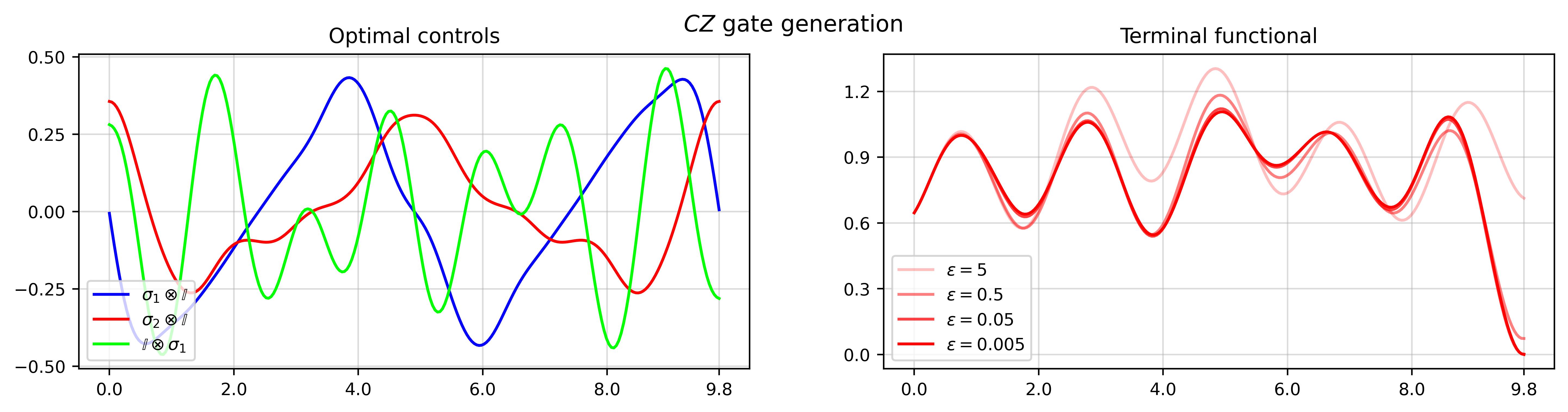}
    \subcaption{}
    \label{fig:BFCS_2b}
  \end{subfigure}
  \caption{Results of the gate generation for (\ref{fig:BFCS_2a}) CNOT via the system \eqref{BFCS_subsection_2qgates} with $ \omega_{1} = 3 $, $ \omega_{2} = 4 $, $ \beta = 1.25 $, $ \alpha = 1 $ and $ T = 4.75 $; (\ref{fig:BFCS_2b}) CZ with $ \omega = 2 $, $ \beta_{1} = 0.5 $, $ \beta_{2} = 0.75 $, $ \alpha = 1 $ and $ T = 9.8 $. The plots for the optimal controls are given for $\varepsilon = 0.005$.}
  \label{fig:BFCS_two_qubit_pictures}
\end{figure}
\begin{table}[htbp!]
  \centering
  \begin{tabular}{|c|c|c|c|c|c|}
    \hline
    \multirow{2}{*}{\textcolor{black}{\textbf{Gate} $ \boldsymbol{U_{\operatorname{gate}}} $}}
    & \multirow{2}{*}{\textcolor{black}{$\boldsymbol{J_{\mathrm{term}}}$ \textbf{/} $\boldsymbol{F^{(\operatorname{bound})}_{\mathrm{AGF}}}$}}
    & \multicolumn{4}{c|}{\textcolor{black}{\textbf{Scaling parameter} $\boldsymbol{\varepsilon}$}} \\
    \cline{3-6}
    & & $5 \cdot 10^{0}$
      & $5 \cdot 10^{-1}$
      & $5 \cdot 10^{-2}$
      & $5 \cdot 10^{-3}$ \\
    \hline
    \multirow{2}{*}{$e^{i \frac{\pi}{4}}\text{CNOT}$}
    & $\textcolor{black}{J_{\mathrm{term}}}$
    & $0.90847$ & $0.20116$ & $0.00501$ & $0.00006$ \\
    \cline{2-6}
    & \textcolor{black}{$F^{(\operatorname{bound})}_{\mathrm{AGF}}$}
    & \textcolor{black}{$0.20670$}
    & \textcolor{black}{$0.71052$}
    & \textcolor{black}{$0.99200$}
    & \textcolor{black}{$0.99990$} \\
    \hline
    \multirow{2}{*}{$e^{i \frac{\pi}{4}}\text{CZ}$}
    & $\textcolor{black}{J_{\mathrm{term}}}$
    & $0.71316$ & $0.07314$ & $0.00189$ & $0.00002$ \\
    \cline{2-6}
    & \textcolor{black}{$F^{(\operatorname{bound})}_{\mathrm{AGF}}$}
    & \textcolor{black}{$0.26582$}
    & \textcolor{black}{$0.88726$}
    & \textcolor{black}{$0.99698$}
    & \textcolor{black}{$0.99997$} \\
    \hline
  \end{tabular}
  \caption{{\color{black} The values of the terminal functional $J_{\mathrm{term}}$ and the corresponding lower bounds $F^{(\operatorname{bound})}_{\mathrm{AGF}}$ on the average gate fidelity for the two-qubit experiments in  Figure~\ref{fig:BFCS_two_qubit_pictures}. Both the initial and final BVP solver mesh sizes are $250$~nodes for all of the two-qubit experiments.}}  \label{tab:BFCS_2q}
\end{table}

For gates (\ref{gates2}), the components $u_{\text{gate}}^{(0)}\in\mathbb{C}$ and the non-zero components of their Bloch vectors in $\mathbb{C}^{15}$   (with the correction phase factor $ e^{i \frac{\pi}{4} }$) have the forms
\begin{equation}
    \begin{split}
        &u_{\text{CNOT}}^{(0)} =  \frac{1}{2 \sqrt{2}} (1 + i), \quad  u_{\text{CNOT}}^{(11)} = \frac{1}{2} (1+i), \quad u_{\text{CNOT}}^{(14)} =  \frac{1}{2\sqrt{3}} (1 + i), \quad u_{\text{CNOT}}^{(15)} =  \frac{1}{2 \sqrt{6}} (1 + i), \\[0.2cm]
        & u_{\text{CZ}}^{(0)} = \frac{1}{2 \sqrt{2}} (1 + i) , \quad  u_{\text{CZ}}^{(15)} =  \frac{\sqrt{3}}{2 \sqrt{2}} (1 + i).
    \end{split}
\end{equation}

In the numerical experiments, the optimal implementation of gate CNOT is considered within the unitary evolution of a two-qubit system described by the Hamiltonian (\ref{BFCS_two_qubit_hamiltonian}) with isotropic interaction $ \beta_{1} = \beta_{2} =: \beta $ and unequal frequency values $ \omega_{1} \neq \omega_{2} $, while the optimal implementation of gate CZ -- with an anisotropic interaction $ \beta_{1} \neq \beta_{2} $ and equal frequencies $ \omega_{1} = \omega_{2} =: \omega $. Corresponding numerical results are presented in Figures \ref{fig:BFCS_two_qubit_pictures} and Table \ref{tab:BFCS_2q}.\\[0.5cm]

\subsection{Three-qubit gate }\label{BFCS_subsection_3qgates}
As an example in a three-qubit case, we consider the implementation of the Toffoli gate (CCNOT)
\begin{equation}
    \begin{split}
        \text{Toffoli} &= |0\rangle \langle 0| \otimes \mathbb{I}_{4} + |1\rangle \langle1| \otimes \text{CNOT} = \\[0.2cm]
        &= |00\rangle \langle 00| \otimes \mathbb{I}_{2} + |01\rangle \langle 01| \otimes \mathbb{I}_{2} + |10\rangle \langle 10| \otimes \mathbb{I}_{2} + |11\rangle \langle 11| \otimes \text{NOT},
    \end{split}
\end{equation}
(where $ |\alpha \beta \rangle := |\alpha \rangle \otimes |\beta \rangle $) within the unitary evolution of the three-qubit quantum system with the Hamiltonian  
\begin{equation}\label{BFCS_three_qubit_hamiltonian}
    \begin{split}
        H &= \frac{\omega_{1}}{2} \, \sigma_{3} \otimes \mathbb{I} \otimes \mathbb{I} + \frac{\omega_{2}}{2} \, \mathbb{I} \otimes \sigma_{3} \otimes \mathbb{I} + \frac{\omega_{3}}{2} \, \mathbb{I} \otimes \mathbb{I} \otimes \sigma_{3} + \sum\limits_{q \in \{2, \, 3\}} \left(\beta_{12}^{q} \, \sigma_{q} \otimes \sigma_{q} \otimes \mathbb{I} + \beta_{23}^{q} \, \mathbb{I} \otimes \sigma_{q} \otimes \sigma_{q} \right) + \\[0.2cm]
        & + \nu_{1}(t) \, \sigma_{1} \otimes \mathbb{I} \otimes \mathbb{I} + \nu_{2}(t) \, \mathbb{I} \otimes  \sigma_{1} \otimes \mathbb{I} + \nu_{3}(t) \, \sigma_{2} \otimes \mathbb{I} \otimes \mathbb{I} + \nu_{4}(t) \, \mathbb{I} \otimes \mathbb{I} \otimes \sigma_{2}, 
    \end{split}
\end{equation}
which inherits the structure of the Hamiltonians \eqref{BFCS_one_qubit_hamiltonian} and \eqref{BFCS_two_qubit_hamiltonian} in the sense of interaction couplings between qubits forming a chain: the first qubit with the second qubit and the second with the third. For controlling the three-qubit system, we use the local actions on qubits by: $ \sigma_{1} $ for the second qubit, $ \sigma_{2} $ for the third one, and both of them for the first.

Since, in a three-qubit case, the Bloch vector $ h $ of the Hamiltonian (\ref{BFCS_three_qubit_hamiltonian}) and the Bloch vector $ u_{\text{gate}} $ of the Toffoli gate have $63$ components each, we do not include here the values of these components. Within the numerical experiments, they are calculated via \eqref{BFCS_X_decomposition_over_basis}, \eqref{BFCS_Gellman_basis} and  \eqref{BFCS_starter}--\eqref{BFCS_finisher}. 

The derived numerical results are presented in Figure \ref{fig:BFCS_three_qubit_pictures} and Table \ref{tab:BFCS_3q}.

\begin{figure}[!htbp]
  \centering
  \begin{subfigure}[t]{\textwidth}
    \centering
    \includegraphics[width=\linewidth]{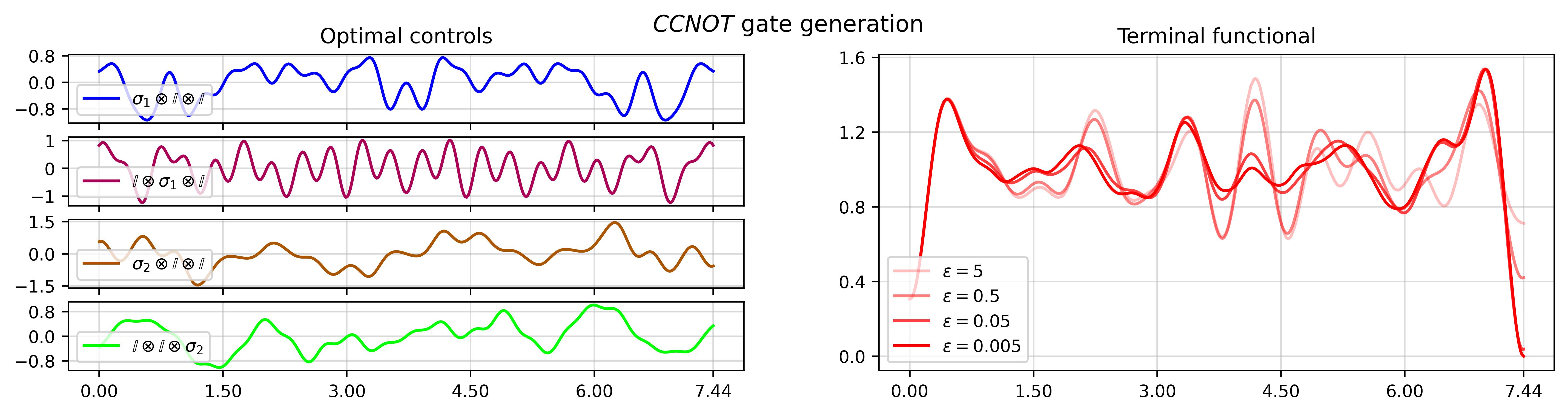}
  \end{subfigure}
  \caption{Results of the CCNOT gate generation via the system \eqref{BFCS_three_qubit_hamiltonian} with $ \omega_{1} = 1 $, $ \omega_{2} = 2 $, $ \omega_{3} = 3 $, $ \beta_{12}^{y} = 1 $, $ \beta_{12}^{z} = 3 $, $ \beta_{23}^{y} = 5 $, $ \beta_{23}^{z} = 1.5 $ and $ T = 7.44 $. The plots for the optimal controls are given for $\varepsilon = 0.005$.}
  \label{fig:BFCS_three_qubit_pictures}
\end{figure}

\begin{table}[htbp!]
  \centering
  \begin{tabular}{|c|c|c|c|c|c|}
    \hline
    \multirow{2}{*}{\textcolor{black}{\textbf{Gate} $ \boldsymbol{U_{\operatorname{gate}}} $}}
    & \multirow{2}{*}{\textcolor{black}{$\boldsymbol{J_{\mathrm{term}}}$ \textbf{/} $\boldsymbol{F^{(\operatorname{bound})}_{\mathrm{AGF}}}$}}
    & \multicolumn{4}{c|}{\textcolor{black}{\textbf{Scaling parameter} $\boldsymbol{\varepsilon}$}} \\
    \cline{3-6}
    & & $5 \cdot 10^{0}$
      & $5 \cdot 10^{-1}$
      & $5 \cdot 10^{-2}$
      & $5 \cdot 10^{-3}$ \\
    \hline
    \multirow{2}{*}{$e^{i \frac{\pi}{8}}\text{CCNOT}$}
    & $\textcolor{black}{J_{\mathrm{term}}}$
    & $0.7124$ & $0.4207$ & $0.0385$ & $0.0007$ \\
    \cline{2-6}
    & \textcolor{black}{$F^{(\operatorname{bound})}_{\mathrm{AGF}}$}
    & \textcolor{black}{$0.1846$}
    & \textcolor{black}{$0.4094$}
    & \textcolor{black}{$0.9329$}
    & \textcolor{black}{$0.9988$} \\
    \hline
  \end{tabular}
  \caption{{\color{black}The values of the terminal functional $J_{\mathrm{term}}$ and the corresponding lower bounds $F^{(\operatorname{bound})}_{\mathrm{AGF}}$ on the
  average gate fidelity for the three-qubit experiment in Figure~\ref{fig:BFCS_three_qubit_pictures}.
  The initial BVP solver mesh size is $ 100 $~nodes, the final size is $ 397 $.}}
  \label{tab:BFCS_3q}
\end{table}

\newpage
\section{Conclusion}\label{BFCS_section_conclusion}
The present paper addresses the problem on the optimal implementation of $N$-qubit quantum gates for closed quantum systems. Based on the generalized Bloch vectors formalism \cite{26}  for a finite-dimensional quantum system,  extending the Pontryagin principle  \cite{6} to the case of complex variables and using the special scaling in the cost functional (\ref{BFCS_general_functional}),  we: (i) develop (Section~\ref{BFCS_section_general_optimal_gates_realization}) a new optimal model (\ref{BFCS_model_statement}) which has the unified mathematical form,{\color{black} which is standard for the mathematical control theory and applicable} for the implementation of an arbitrary $N$-qubit gate within any closed $N$-qubit system, satisfying the controllability conditions \cite{2,3'};  (ii) present (Theorem~\ref{BFCS_theorem_with_conds}, Section~\ref{BFCS_section_general_optimal_gates_realization}) the necessary conditions for optimality of the solution of the developed optimal model; (iii) validate numerically (Section \ref{BFCS_section_numerical}) the solutions of the new optimal model (\ref{BFCS_model_statement}) for the implementation of a variety of one-qubit, two-qubit and three-qubit quantum gates. 

The developed optimal model allows generating a quantum gate both with and without taking into account its precise global phase. The used cost functional \eqref{BFCS_general_functional} consists of both the terminal part, measuring the proximity of the unitary evolution operator of a closed $ N $-qubit system at a finite time to the target unitary operator of a designed gate, and also, the integral part, describing from the physical point of view the energy costs for  gate implementation, and  used with a sufficiently small scaling factor  \eqref{BFCS_integral_elem} -- this allows us to obtain a convenient numerical method for the optimal control synthesis.

{\color{black} In contrast to \cite{14,18,21,22,19}, the approach, developed in the present article, sharply separates the analytical stage (by finding the necessary optimality conditions \eqref{BFCS_final_equations}, Section~\ref{BFCS_section_general_optimal_gates_realization}) and the numerical stage (Section~\ref{BFCS_section_numerical}) of the optimal problem analysis}.

For conducting the numerical experiments, we have used the SciPy bvp solver \cite{30}, which iteratively builds continuous  optimal controls. We stress that this solver can be replaced by any other suitable numerical tool for the ODEs system \eqref{BFCS_final_equations}. 

The validating numerical tests have been performed for a variety of $N=1,2,3$ quantum gates widely used in applications: (i) one-qubit gates NOT, Hadamard (H), phase (S), and $ \frac{\pi}{8} $ (T); (ii) two-qubit gates like CNOT and CZ; and (iii) the three-qubit Toffoli gate. We stress that the implementation of the Toffoli gate has not been considered within the theoretical optimal models proposed in \cite{14, 16, 18, 21, 22, 19, 19'} and references therein. 

\textcolor{black}{The obtained numerical results presented in Tables~\ref{tab:BFCS_1q},~\ref{tab:BFCS_2q},~\ref{tab:BFCS_3q} clearly indicate that for small values of the scaling parameter $ \varepsilon$ in the cost functional (\ref{BFCS_general_functional}),  the developed optimal model (\ref{BFCS_model_statement}) enables the implementation of one-qubit, two-qubit and three-qubit gates with a high degree of precision, which is evidenced by the small final values of the terminal functional (\ref{BFCS_terminal_functional}) and the corresponding values of the average gate fidelity (\ref{agf}), which are close to 1.} 

\textcolor{black}{The optimal model (\ref{BFCS_model_statement}), developed in this article, is formulated for the implementation of quantum gates through the evolution of closed quantum systems. In a sequel to this article, building on its results and the description  of the open quantum system evolution in terms of the generalized Bloch vectors formalism in \cite{26},  we will present a generalization of the developed approach to the implementation of quantum gates through the evolution of open quantum systems.}

\section{Acknowledgments}

The study was implemented in the framework of the Basic Research Program at
the HSE University (HSE-BR-2025-007). The valuable comments of the anonymous Referee are highly appreciated.

\bigskip

\noindent \textbf{Funding} This research did not receive funding.

\end{document}